\pdfoutput=1
\documentclass[fleqn,usenatbib]{mnras}

\usepackage{newtxtext,newtxmath}

\usepackage[T1]{fontenc}

\DeclareRobustCommand{\VAN}[3]{#2}
\let\VANthebibliography\thebibliography
\def\thebibliography{\DeclareRobustCommand{\VAN}[3]{##3}\VANthebibliography}

\usepackage{graphicx}   
\usepackage{amsmath}    

\usepackage{multirow}
\usepackage{diagbox}
\usepackage
{caption}
\usepackage{subcaption}

\usepackage{orcidlink}
\usepackage{xcolor}

\usepackage{hyperref}
\usepackage{appendix}

\usepackage{tabularx}

\newcommand{\aref}[1]{\hyperref[#1]{Appendix~\autoref{#1}}}
\defcitealias{vijayan24}{QED I}

\title[SMPBHs explain LRDs]{Sub-Eddington accreting supermassive primordial black holes explain Little Red Dots}

\author[Huang et al.]{Hai-Long Huang \orcidlink{0009-0003-1218-2569}$^{1, 2}$\thanks{E-mail: huanghailong18@mails.ucas.ac.cn},
Jun-Qian Jiang\orcidlink{0000-0003-0957-3633}$^{2}$\thanks{E-mail:  jiangjunqian21@mails.ucas.ac.cn, jiangjq2000@gmail.com},
Jibin He$^{3}$\thanks{E-mail: 20242701016@stu.cqu.edu.cn},
Yu-Tong Wang$^{1,2}$\thanks{E-mails:wangyutong@ucas.ac.cn}
Yun-Song Piao$^{1, 2, 4, 5}$\thanks{E-mail: yspiao@ucas.ac.cn}
\\
$^{1}$School of Fundamental Physics and Mathematical Sciences, Hangzhou Institute for Advanced Study, UCAS, Hangzhou 310024, China\\
$^{2}$School of Physical Sciences, University of Chinese Academy of Sciences, Beijing 100049, China\\
$^{3}$Department of Physics and Chongqing Key Laboratory for Strongly Coupled Physics, Chongqing University, Chongqing 401331, P. R. China\\
$^{4}$International Center for Theoretical Physics Asia-Pacific, Beijing/Hangzhou, China\\
$^{5}$Institute of Theoretical Physics, Chinese Academy of Sciences, P.O. Box 2735, Beijing 100190, China
}

\date{Accepted XXX. Received YYY; in original form ZZZ}

\pubyear{2024}

\begin{document}
\label{firstpage}
\pagerange{\pageref{firstpage}--\pageref{lastpage}}
\maketitle

\begin{abstract}
The James Webb Space Telescope (JWST) has uncovered an abundant
population of compact, extremely red, and X-ray weak objects at
$z\gtrsim4$, knows as ``Little Red Dots" (LRDs). These objects
exhibit spectral energy distributions that resemble both active
galactic nuclei (AGN) and stellar population templates. However,
whether dominated by AGN activity or compact star formation, the
high redshifts and masses/luminosities of LRDs, coupled with their
significant abundance, present potential challenges to the
standard $\Lambda$CDM model. In this work, we proposes a novel
cosmic interpretation of this anomaly, suggesting that these LRDs
are likely massive galaxies seeded by supermassive primordial
black holes (SMPBHs) came into being in the very early universe.
We analyze 434 known LRDs from the 0.54 ${\rm deg}^2$ COSMOS-Web
survey and test the hypothesis that they originated from SMPBHs
assuming sub-Eddington accretion. According to our result, SMPBHs
actually could lead to the existence of more LRDs, even at higher
redshifts ($z>8$).


\end{abstract}

\begin{keywords}
Early universe --- Galaxy formation --- Supermassive black holes
\end{keywords}


\section{Introduction} \label{sec:intro}

The James Webb Space Telescope (JWST) offers an unprecedented
glimpse into the early universe, extending the boundaries of
detectable supermassive black holes (SMBHs) and their host
galaxies in both redshift and mass \citep{2023ApJ...953L..29L,
2023A&A...677A.145U,
2023ApJ...955L..24G,2023ApJ...957L...7K,2023arXiv231203589U,
2023arXiv230512492M,2024NatAs...8..126B,
2024ApJ...960L...1N,2024ApJ...965L..21K}. Notably, JWST has
revealed a previously unknown population of dust-reddened galaxies
at redshifts $z\gtrsim4$, so-called ``Little Red Dots" (LRDs)
\citep{2023ApJ...959...39H,2023ApJ...954L...4K,2023arXiv230801230M,
2023arXiv231203065K,2024ApJ...963..129M,2024arXiv240810305G,
2024arXiv240715915P,2024arXiv240913441F,2024arXiv241006257M},
posing new challenges to our understanding of early galaxy
formation and SMBH growth. They have very red near-infrared colors
and are very compact, with a median effective radius of
$\sim150$pc, while some are even smaller with $<35$pc
\citep{2023ApJ...955L..12B,
2023ApJ...952..142F,2020ApJ...895...95P}. Additionally, these
sources are abundant compared to the previously known population
of high-$z$ active galactic nuclei (AGN). Their number density is
$10^{-4}-10^{-5} {\rm Mpc}^{-3}{\rm mag}^{-1}$ at $z\sim5$, and is
$10-100$ times more numerous than the faint end of quasar
luminosity function \citep{2024ApJ...964...39G}.

Their physical interpretations bifurcated into two paths. The
continuum emission in LRDs could be either dominated by the direct
thermal emission from the accretion disk of AGN
\citep{2023arXiv230607320L,2023arXiv230605448M}, or from a
population of young stars associated with vigorous star formation
\citep{2023Natur.616..266L,2023ApJ...956...61A,2024ApJ...968...34W,
2024ApJ...968....4P,2024arXiv240811890A}. However, if LRDs are
dominated by star formation, they would dominate the high-mass end
of the stellar mass function, and have central stellar mass
densities significantly greater than the maximum observed at lower
redshift \citep{2024arXiv240610341A}. And based on their compact
sizes, most of them must approach or exceed the maximal stellar
mass surface densities observed in the local universe. By
contrast, if LRDs are AGN-dominated, the high SMBH masses imply
that the black hole-to-stellar mass ratio of LRDs is much higher
than that in the local universe \citep{2024ApJ...969L..18A}, and
an overbundance of SMBHs, about two orders of magnitude larger
than was expected from pre-JWST observations of the bolometric
quasar luminosity function
\citep{2024ApJ...968...38K,2024arXiv240403576K,2024arXiv240705094C}.

There is actually clear evidence for AGN in the LRDs due to
ubiquitous broad lines, and also evidence for evolved stellar
populations with clear Balmer breaks \citep{2024ApJ...964...39G,
2024ApJ...969L..13W}. Therefore, LRDs could be a heterogeneous
galaxy population, in which both AGN and star formation contribute
to their fluxes to different degree. However,
\citet{2024arXiv240610329D} found that, independent of the
specific AGN fraction adopted, the LRDs' black holes are
significantly overmassive relative to their host galaxies compared
to the local $M_{\rm BH}-M_*$ relation. Moreover, the first
spectroscopic follow-ups showed a significant presence of broad
emission lines in their spectra, suggesting the existence of AGN
in their cores with a high $M_{\rm BH}/M_*$ ratio
\citep{2023ApJ...954L...4K,2023arXiv230805735F,2024ApJ...964...39G}.

\begin{figure*}
     \centering
    $$
    \begin{array}{c}
        \includegraphics[width=2\columnwidth]{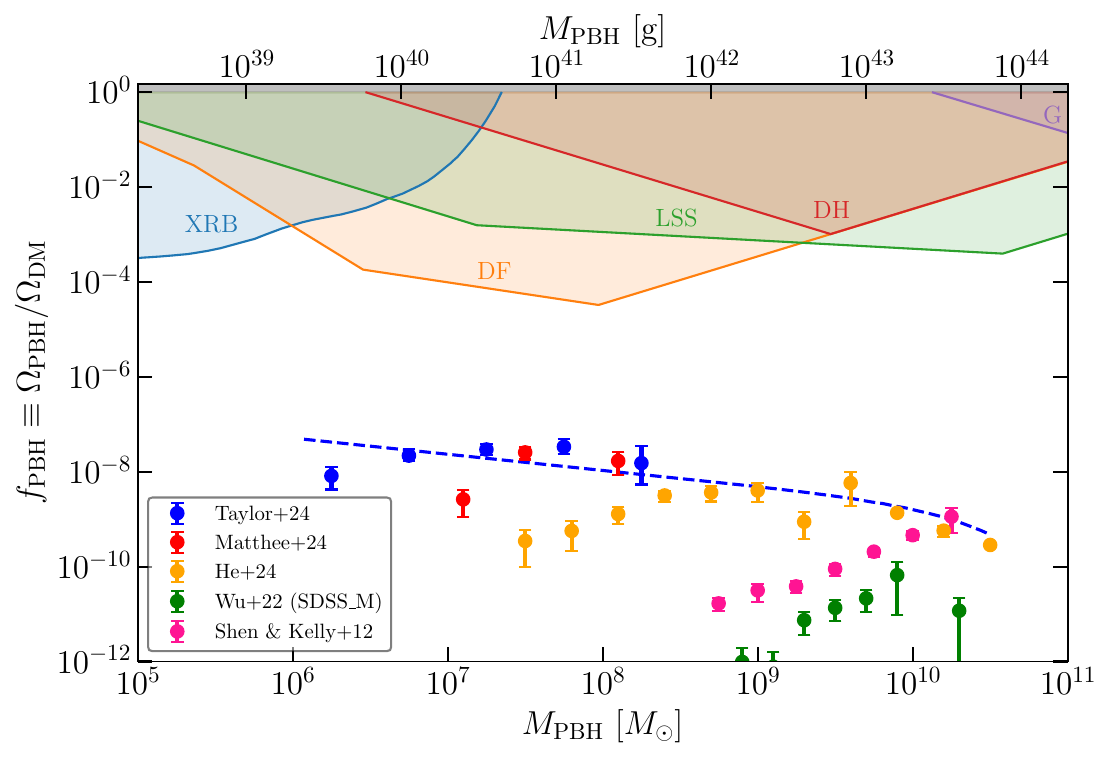}
    \end{array}
    $$
    \caption{Comparison between the derived SMBH mass
    function and constraints on the fraction of SMPBHs in DM.
The observational constraints are from heating of stars in the
Galactic disk (DH), infalling of halo objects due to
    dynamical friction (DF), tidal disruption of galaxies (G), accretion limits come
    from X-ray binaries (XRB), and large-scale structure constraints (LSS) \citep{2021RPPh...84k6902C}.
    Green points show the SMBH mass function at $z\sim6$ from \citep{2022MNRAS.517.2659W},
    and pink points show the SMBH mass function from \citep{2012ApJ...746..169S},
    both derived from SDSS data. Yellow points show the SMBH mass function
    at $3.5<z<4.25$ derived from ground-based SDSS+HSC data in \citep{2024ApJ...962..152H}.
    Red points show the SMBH mass function derived from NIRCam WFSS data in
    \citep{2024ApJ...963..129M}. Blue points show the SMBH mass function at
    $3.5<z<6.0$ derived from data from the CEERS and RUBIES surveys (including
    LRDs) in \citep{2024arXiv240906772T}. Dashed blue curve shows a Schecter
    fit to the data in \citep{2024arXiv240906772T} from $6.5<\log M_{\rm BH}
    <8.5$ and the data in \citep{2024ApJ...962..152H} at $\log M_{\rm BH}>8.5$. }
    \label{fig:smpbh}
\end{figure*}

The existence of such overmassive black holes in the high-redshift
Universe seems to provide the strongest evidence yet of heavy
black hole seeding occurring during the cosmic dark ages. However,
we propose that these SMBHs could have come into being in the very
early universe as supermassive primordial black holes (SMPBHs).
Recently, the cosmological implications of PBHs
\citep{1974MNRAS.168..399C,Zeldovich:1967lct,1971MNRAS.152...75H}
have been intensively studied, see reviews by e.g.
\citet{2016PhRvD..94h3504C,2017JPhCS.840a2032G,
2018CQGra..35f3001S,2020ARNPS..70..355C,
2021RPPh...84k6902C,2021JPhG...48d3001G,2021arXiv211002821C,
2022arXiv221105767E,2024arXiv240406151P,2024CQGra..41n3001D}. It
is important to note that the required abundance of SMPBHs to
explain the energy density of SMBHs can be compatible with current
observations, see \autoref{fig:smpbh} (see also
\autoref{appendixB}). These SMPBHs may originate from the direct
collapse of large-amplitude non-Gaussian primordial perturbations
after horizon entry
\citep{2024JCAP...09..012B,2018PhRvD..97d3525N,2021PhRvD.103f3519A,
2023EL....14249001G,2024JCAP...04..021H}, or through other
non-primordial-perturbation-like mechanisms e.g.
\citet{2023arXiv230617577H,2024PhRvD.110b3501H}. They can assemble
galaxies through the seed effect \footnote{The seed effect relates
to the gravitational influence of individual black holes, while
the Poisson effect accounts for the collective $\sqrt{N}$
fluctuations in the number of black holes. See e.g.
\citet{1966RSPSA.290..177H,1975A&A....38....5M,
1983ApJ...268....1C,1984MNRAS.206..801C,2018MNRAS.478.3756C,
2024arXiv240715781H,2024arXiv240707162D} for more detail.} without
requiring significant growth themselves.


The rest of this paper is structured as follows. In
\autoref{sec:data}, we describe the COSMOS-Web survey and our data
selection.
In \autoref{sec:methods}, we
provide a comprehensive overview of the SMPBH model and present a
detailed description of the data analysis methodology employed in this study.
Lastly, we summarise our conclusions and offer a discussion of the implications
in \autoref{sec:results}.
Throughout this paper the values
of cosmological parameters are set in light of the Planck results
\citep{2020A&A...641A...6P}.


\section{Data} \label{sec:data}

We use a sample of 434 $5\lesssim z\lesssim8$ LRDs selected from
the 0.54 ${\rm deg}^2$ COSMOS-Web JWST survey, as reported in
\citet{2024arXiv240610341A}. COSMOS-Web is a major JWST Cycle 1
program that images a contiguous 0.54 ${\rm deg}^2$ area using
NIRCam and a 0.19 ${\rm deg}^2$ area with MIRI
\citep{2023ApJ...954...31C} within the COSMOS field
\citep{2007ApJS..172....1S}. For further details, refer to the
original Ref.\citep{2024arXiv240610341A}. The large sample size,
which nearly doubles the number of LRDs reported in
\citet{2024ApJ...968...38K,2024arXiv240403576K}, and the extensive
on-sky area help minimize the impact of cosmic variance. This
allows us to investigate whether SMPBHs could theoretically serve
as seeds for galaxies with overmassive central black holes.

We present the properties of the partial COSMOS-Web LRDs in
\autoref{tab:i}. These properties were derived based on two extreme
scenarios: QSO models and galaxy models, where the LRD population is
primarily composed of dust-reddened AGN or compact/dusty starbursts.
In these models, it is assumed that the continuum emission in LRDs is
dominated by flux from the accretion disk of AGN or from a population
of young stars, respectively. Ideally, we would fit each object with
a composite spectral energy distribution (SED) model that decomposes both
components, as done in \citet{2024arXiv240610329D}. However, as noted in
\citet{2024arXiv240610341A}, the limited available information makes such an
analysis prone to overfitting. Therefore, we use the results from the two
models to estimate the corresponding SMBH mass and stellar mass, respectively.

\begin{table}
\centering
\begin{tabular}{c|ccc|cc}
\hline
\multirow{3}{*}{ID} & \multicolumn{3}{c|}{QSO models} & \multicolumn{2}{c}{galaxy models}\\
\cline{2-4} \cline{5-6} & $z_{\rm qso}$ & $\log L_{\rm bol}$ & $\log M_{\rm BH}$ & $z_{\rm gal}$  & $\log M_*$  \\
 &  & [erg ${\rm s}^{-1}$] & [$M_\odot$] &  & [$M_\odot$]\\ 
\hline
1491 & $6.2_{-0.5}^{+1.2}$ & $46.2_{-0.3}^{+0.4}$ & $10.1_{-0.3}^{+0.4}$ & 
$6.1_{-0.7}^{+1.2}$ & $10.7_{-0.3}^{+0.3}$ 
\\
3962 & $7.3_{-1.8}^{+1.9}$ & $46.2_{-0.4}^{+0.3}$ & $10.1_{-0.4}^{+0.3}$ & 
$7.0_{-2.3}^{+2.1}$ & $10.4_{-0.2}^{+0.3}$ 
\\
4056 & $6.0_{-0.5}^{+1.0}$ & $45.8_{-0.2}^{+0.2}$ & $9.7_{-0.2}^{+0.2}$ & 
$6.2_{-0.7}^{+2.2}$ & $9.4_{-0.3}^{+0.3}$
\\
8203 & $8.3_{-0.4}^{+0.4}$ & $46.2_{-0.3}^{+0.2}$ & $10.1_{-0.3}^{+0.2}$ & 
$5.3_{-0.2}^{+0.2}$ & $8.7_{-0.1}^{+0.1}$ 
\\
9430 & $7.3_{-1.2}^{+1.0}$ & $45.9_{-0.3}^{+0.3}$ & $9.8_{-0.3}^{+0.3}$ & 
$7.8_{-0.6}^{+0.6}$ & $9.8_{-0.3}^{+0.2}$ 
\\
9680 & $7.5_{-1.4}^{+0.7}$ & $46.3_{-0.3}^{+0.3}$ & $10.2_{-0.3}^{+0.3}$ & 
$5.8_{-0.4}^{+1.6}$ & $9.3_{-0.2}^{+0.2}$ 
\\
10023 & $6.0_{-0.4}^{+1.0}$ & $45.3_{-0.2}^{+0.2}$ & $9.2_{-0.2}^{+0.2}$ & 
$6.0_{-0.6}^{+2.3}$ & $8.9_{-0.2}^{+0.3}$ 
\\
10511 & $5.9_{-0.3}^{+0.4}$ & $45.6_{-0.1}^{+0.1}$ & $9.5_{-0.1}^{+0.1}$ & 
$5.5_{-0.3}^{+0.6}$ & $8.9_{-0.1}^{+0.2}$ 
\\
11737 & $7.8_{-1.8}^{+1.5}$ & $46.5_{-0.4}^{+0.5}$ & $10.4_{-0.4}^{+0.5}$ & 
$7.5_{-2.3}^{+1.7}$ & $10.7_{-0.4}^{+0.7}$
\\
13618 & $6.0_{-0.5}^{+0.6}$ & $45.4_{-0.2}^{+0.2}$ & $9.3_{-0.2}^{+0.2}$ & 
$6.2_{-0.8}^{+2.1}$ & $9.1_{-0.3}^{+0.5}$ 
\\
14599 & $5.9_{-0.4}^{+0.4}$ & $45.5_{-0.2}^{+0.2}$ & $9.4_{-0.2}^{+0.2}$ & 
$5.6_{-0.4}^{+2.1}$ & $8.9_{-0.2}^{+0.3}$ 
\\
15495 & $8.1_{-0.7}^{+0.5}$ & $45.9_{-0.2}^{+0.2}$ & $9.8_{-0.2}^{+0.2}$ & 
$5.5_{-0.3}^{+0.5}$ & $9.1_{-0.1}^{+0.2}$ 
\\
15654 & $5.7_{-0.3}^{+0.4}$ & $46.2_{-0.2}^{+0.2}$ & $10.0_{-0.2}^{+0.2}$ & 
$5.6_{-0.3}^{+0.6}$ & $9.8_{-0.2}^{+0.2}$ 
\\
15961 & $6.2_{-0.5}^{+0.9}$ & $46.3_{-0.3}^{+0.4}$ & $10.1_{-0.3}^{+0.4}$ & 
$6.4_{-0.6}^{+1.2}$ & $10.9_{-0.4}^{+0.4}$
\\
16108 & $6.0_{-0.4}^{+1.3}$ & $45.8_{-0.2}^{+0.2}$ & $9.7_{-0.2}^{+0.2}$ & 
$5.6_{-0.4}^{+0.5}$ & $9.2_{-0.2}^{+0.2}$ 
\\
16692 & $6.6_{-1.3}^{+2.0}$ & $45.5_{-0.4}^{+0.4}$ & $9.4_{-0.4}^{+0.4}$ & 
$6.5_{-1.5}^{+2.1}$ & $9.7_{-0.4}^{+0.5}$ 
\\
17632 & $7.7_{-1.5}^{+1.1}$ & $46.0_{-0.4}^{+0.4}$ & $9.9_{-0.4}^{+0.4}$ & 
$6.9_{-1.5}^{+1.8}$ & $10.3_{-0.5}^{+0.6}$ 
\\
18065 & $5.9_{-0.5}^{+1.9}$ & $45.8_{-0.4}^{+0.3}$ & $9.7_{-0.4}^{+0.3}$ & 
$5.2_{-0.7}^{+1.9}$ & $10.0_{-0.4}^{+0.4}$ 
\\
19068 & $6.5_{-1.4}^{+2.1}$ & $45.2_{-0.4}^{+0.3}$ & $9.1_{-0.4}^{+0.3}$ & 
$6.7_{-1.6}^{+2.0}$ & $9.4_{-0.3}^{+0.4}$ 
\\
26156 & $7.3_{-1.5}^{+1.8}$ & $46.0_{-0.5}^{+0.6}$ & $9.9_{-0.5}^{+0.6}$ & 
$7.0_{-1.8}^{+1.9}$ & $10.3_{-0.5}^{+0.7}$ 
\\
27062 & $6.1_{-0.9}^{+1.8}$ & $45.3_{-0.4}^{+0.4}$ & $9.2_{-0.4}^{+0.4}$ & 
$5.7_{-1.1}^{+2.2}$ & $9.4_{-0.4}^{+0.5}$ 
\\
27100 & $6.9_{-1.2}^{+1.7}$ & $45.8_{-0.4}^{+0.4}$ & $9.7_{-0.4}^{+0.4}$ & 
$6.7_{-1.6}^{+1.7}$ & $10.2_{-0.5}^{+0.6}$ 
\\
29753 & $6.0_{-0.6}^{+1.6}$ & $45.6_{-0.3}^{+0.2}$ & $9.5_{-0.3}^{+0.2}$ & 
$6.8_{-1.3}^{+1.5}$ & $9.6_{-0.3}^{+0.3}$ 
\\
36933 & $6.4_{-0.9}^{+1.6}$ & $45.3_{-0.3}^{+0.3}$ & $9.2_{-0.3}^{+0.3}$ & 
$7.2_{-1.5}^{+1.3}$ & $9.3_{-0.3}^{+0.2}$ 
\\
37490 & $6.5_{-1.2}^{+1.8}$ & $45.6_{-0.3}^{+0.3}$ & $9.4_{-0.3}^{+0.3}$ & 
$6.4_{-1.4}^{+1.9}$ & $9.5_{-0.3}^{+0.3}$ 
\\
39192 & $6.7_{-1.1}^{+2.0}$ & $45.7_{-0.4}^{+0.4}$ & $9.5_{-0.4}^{+0.4}$ & 
$6.4_{-1.5}^{+2.0}$ & $9.9_{-0.4}^{+0.6}$ 
\\
42832 & $7.6_{-1.5}^{+1.7}$ & $46.0_{-0.4}^{+0.5}$ & $9.9_{-0.4}^{+0.5}$ & 
$7.9_{-2.2}^{+1.4}$ & $10.4_{-0.6}^{+1.0}$ 
\\
42963 & $8.1_{-1.9}^{+1.3}$ & $46.1_{-0.5}^{+0.6}$ & $10.0_{-0.5}^{+0.6}$ & 
$8.1_{-2.3}^{+1.0}$ & $10.5_{-0.6}^{+0.9}$
\\
44171 & $7.2_{-1.6}^{+2.0}$ & $45.9_{-0.4}^{+0.4}$ & $9.8_{-0.4}^{+0.4}$ & 
$6.9_{-2.0}^{+1.9}$ & $10.1_{-0.3}^{+0.4}$ 
\\
46758 & $5.7_{-0.2}^{+0.2}$ & $46.3_{-0.1}^{+0.1}$ & $10.2_{-0.1}^{+0.1}$ & 
$5.2_{-0.2}^{+0.2}$ & $9.6_{-0.1}^{+0.1}$ 
\\
47467 & $5.4_{-0.2}^{+0.3}$ & $46.2_{-0.2}^{+0.2}$ & $10.1_{-0.2}^{+0.2}$ & 
$5.4_{-0.3}^{+0.4}$ & $9.9_{-0.2}^{+0.2}$ 
\\
47745 & $5.9_{-0.4}^{+0.5}$ & $45.5_{-0.2}^{+0.2}$ & $9.4_{-0.2}^{+0.2}$ & 
$5.8_{-0.5}^{+2.3}$ & $9.1_{-0.3}^{+0.4}$ 
\\
48256 & $6.3_{-0.4}^{+1.8}$ & $46.3_{-0.2}^{+0.1}$ & $10.2_{-0.2}^{+0.1}$ & 
$5.7_{-0.4}^{+0.7}$ & $9.6_{-0.2}^{+0.2}$ 
\\
\hline
\end{tabular}
\caption{
List of derived physical properties for the partial COSMOS-Web LRDs 
reported in \citet{2024arXiv240610341A} (visit \url{https://github.com/hollisakins/akins24_cw} for the complete dataset,
and we plot it in the $M_{\rm BH}-M_*$ plane in \autoref{fig:massrelation}). 
Columns are as follows: (1) UNCOVER ID; (2) Redshift of the LRDs 
derived from QSO models; (3) Bolometric luminosity from QSO 
models; (4) SMBH mass inferred from the bolometric luminosity, assuming sub-Eddington
accretion with $f_{\rm Edd}=0.01$; (5) Redshhift of the LRDs based on galaxy models;
(6) Stellar mass based on galaxy models.}
\label{tab:i}
\end{table}

\begin{figure*}
     \centering
    $$
    \begin{array}{cc}
    \centering
    \includegraphics[width=1.06\columnwidth]{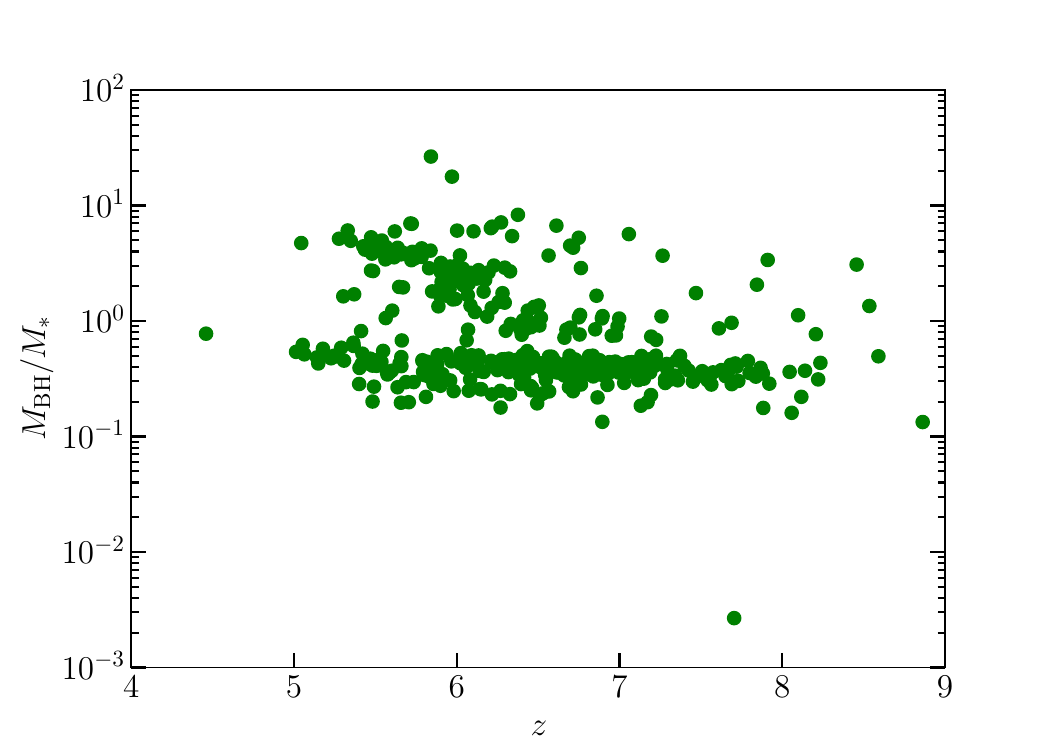} &
    \includegraphics[width=0.94\columnwidth]{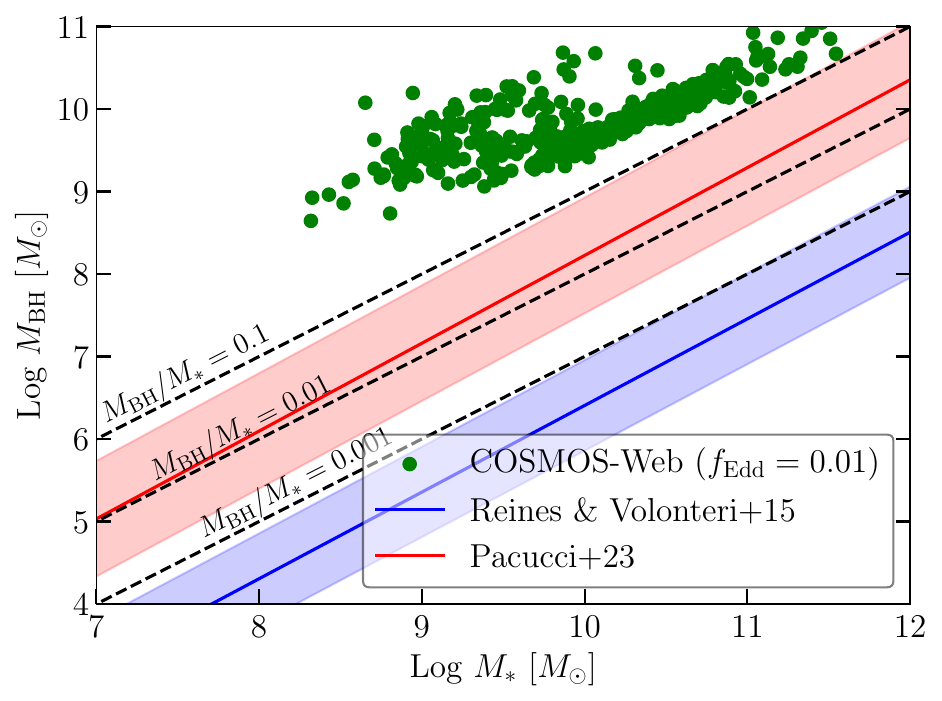}
    \end{array}
    $$
    \caption{\textit{Left panel}: Distribution of the LRDs' black hole-to-star
    mass ratios at various redshifts assuming
    sub-Eddington accretion with $f_{\rm Edd}=0.01$. 
    \textit{Right panel}: The relationship between the SMBH masses
    and the stellar masses of LRDs (green points). The red line represents the high
    redshift
    $M_{\rm BH}-M_*$ scaling relation with its intrinsic scatter \citep{2023ApJ...957L...3P},
    and the blue line represents the local scaling relation from
    \citep{2015ApJ...813...82R}.
    Determining the location of these high-$z$ systems in the $M_{\rm BH}-M_*$
    plane crucially depends on how the contributions from the stellar and the
    AGN components of the SEDs are evaluted.}
    \label{fig:massrelation}
\end{figure*}

The SMBH mass listed in \autoref{tab:i} is inferred from the bolometric
luminosity assuming sub-Eddington accreion.
Specifically, we set the normalized accretion rate \citep{2011ApJ...730....7T}
as follows:
\begin{equation}
    f_{\rm Edd}\equiv\frac{L_{\rm bol}}{L_{\rm Edd}}=0.01,
\end{equation}
where the Eddingtom luminosity is defined by \citep{eddington1926book}
\begin{equation}
    L_{\text{Edd}}=\frac{4\pi GMm_pc}{\sigma_T}\approx1.3
    \times10^{38}\left(\frac{M}{M_\odot}\right)
    ~\textrm{erg}~\textrm{s}^{-1}\mbox{\ ,}
\end{equation}
where
$m_p$ is the proton mass, $c$ is the speed of light and $\sigma_T$
is the Thomson scattering cross-section. However, recent
observations of high-redshift AGN suggest that many may be
dormant, exhibiting sub-Eddington accretion rates
\citep{2023MNRAS.524..176J}. For instance,
\citet{2024arXiv240303872J} report a SMBH with a mass of
approximately $4 \times 10^8M_\odot$, accreting at a rate of only
0.02 times the Eddington limit (i.e. $f_{\rm Edd} = 0.02$) at $z =
6.68$. Other relevant recent studies include e.g.
\citet{2024arXiv240900471W,2024arXiv240903829W,
2024arXiv240502242C,2024arXiv240911457S,2024arXiv240910666N}. In
this context, the mass of a SMBH at a given bolometric luminosity
is estimated to be greater, resulting in a larger black
hole-to-stellar mass ratio. This is illustrated in
\autoref{fig:massrelation}.


\section{Methods} \label{sec:methods}

\subsection{The SMPBH model} \label{subsec:model}

The seeding of early galaxies by SMPBHs was investigated by
\citet{2024arXiv241005891H}. The method is outlined as follows.

After formed in the radiation-dominated era, the SMPBHs with mass
$M_i$ can bind to a dark matter (DM) halo in the subsequent
matter-dominated era, with a mass given by
\citep{2018MNRAS.478.3756C}
\begin{equation}
    M_h=\frac{1+z_{\rm eq}}{1+z}M_i
\end{equation}
at redshift $z$. Its pregalactic growth is considered negligible
\citep{2024arXiv240715781H}.
The binding mass is derived from the fact that PBH
fluctuations grow as $(1+z)^{-1}$ during the matter-dominated
era, but this growth breaks down in the non-linear regime for
$z<z_{\rm cut}$.
Then the mass growth rate of DM halos after $z_{\rm cut}$
with the mass of $M_h(z_{\rm cut})=(1+z_{\rm eq})/(1+z_{\rm cut})M_i$
is characterized with an
analytical function of the form \citep{2010MNRAS.406.2267F,
2022ApJ...938L..10I}
\begin{equation}  \label{eq:dotMh}
    \dot{M}_h\simeq  ~46.1{\rm~M_\odot}~{\rm yr}^{-1}\left(\frac{M_h}
    {10^{12}M_\odot}\right)^{1.1}
    (1+1.11z)\sqrt{\Omega_m(1+z)^3+\Omega_\Lambda}.
\end{equation}
Its analytical solutions can be found in \autoref{appendixA}.
If we allow star formation to take place at $z_{\rm cut}$,
we have
\begin{align} \label{eq:Mstar}
    M_*=
\begin{cases}
    0  \qquad  \qquad {\rm if}  \qquad z>z_{\rm cut},  \\
    \epsilon_* f_bM_h \quad  \ \, {\rm otherwise},
\end{cases}
\end{align}
where $f_b=\Omega_b/\Omega_m$ is the cosmic average fraction
of baryons in matter and $\epsilon_*$ is the star formation
efficiency that is assumed to be constant here.

\begin{figure*}
     \centering
    $$
    \begin{array}{cc}
    \centering
    \includegraphics[width=\columnwidth]{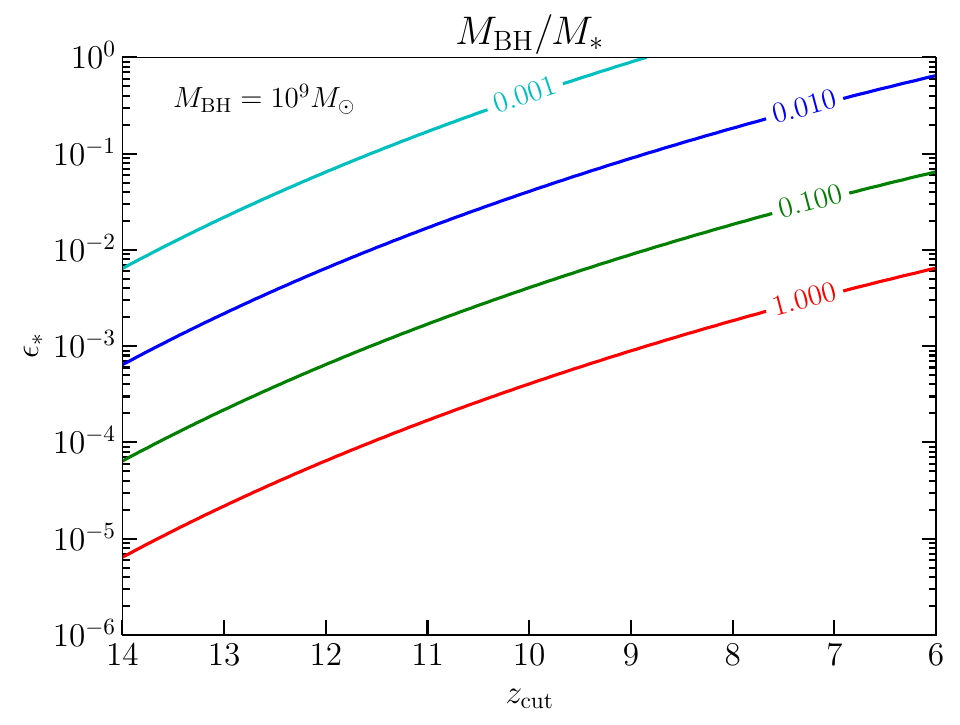} &
    \includegraphics[width=\columnwidth]{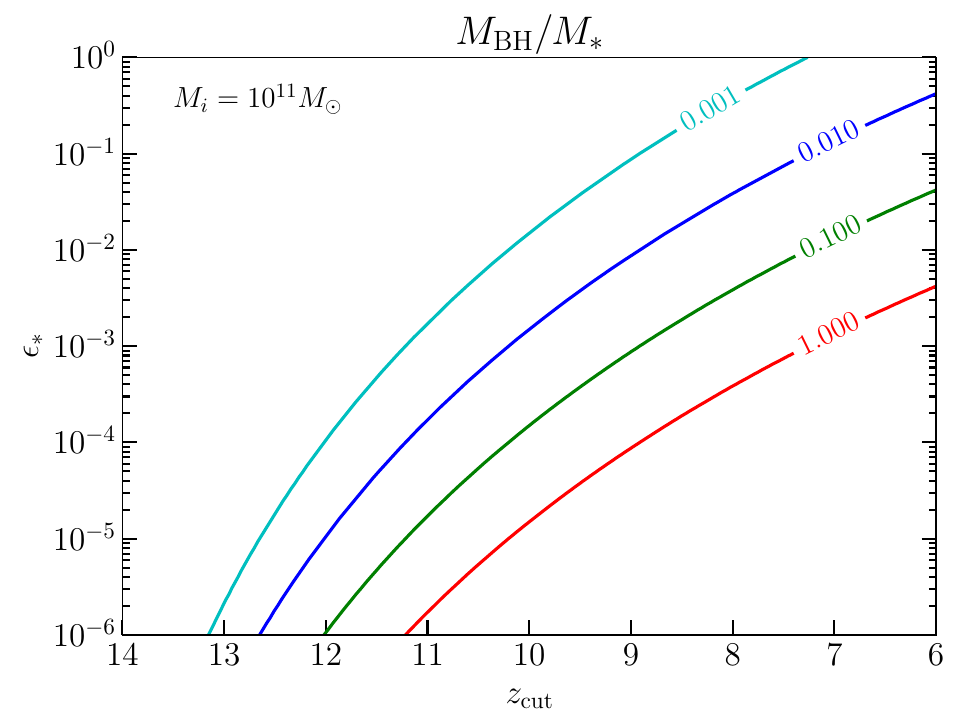}
    \end{array}
    $$
    \caption{\textit{Left panel}: Contour of the ratio of the SMBH mass
    to the stellar mass $z=5$ in the parameter space
    $(z_{\rm cut}, \epsilon_*)$
    in the case of $M_i=M_{\rm BH}=10^{9}M_\odot$.
    \textit{Right panel}: Same as the left panel with
    the different parameters $M_{\rm BH}=10^{11}M_\odot$.}
    \label{fig:parameterspace}
\end{figure*}

On the other hand, the time evolution of the SMBH mass
at $z<z_{\rm cut}$ can be expressed as
\begin{equation}
    M_{\rm BH}(t)=M_{\rm cut}\exp\left[\frac{4\pi Gm_pf_{\rm Edd}}{c\sigma_T}\frac
    {1-\epsilon}{\epsilon}(t-t_{\rm cut})\right].
\end{equation}
Here, $M_{\rm cut}=M(z_{\rm cut})\simeq M_i$ is the SMPBH mass at $z_{\rm cut}$
and $\epsilon$ is the radiative efficiency.
The cosmic time $t$ and redshift $z$ are related by
\begin{equation}
    t(z)=\int_z^{\infty}\frac{{\rm d}z'}{H(z')(1+z')},
\end{equation}
where $H(z)$ is the Hubble parameter. Here, we consider the
sub-Eddington accretion with $f_{\rm Edd}\ll 1$, and thus have
$M_{\rm BH}(z)=M_{\rm cut}=M_i$ approximately. Therefore, in the
SMPBH model with negligible accretion, the stellar mass $M_*(z)$
of a galaxy seeded by an SMPBH can be straightforwardly obtained
using \autoref{eq:Mstar}. As an example, we plot the contour of
the ratio of the SMBH mass to the stellar mass at $z=5$ in the
parameter space $(z_{\rm cut}, \epsilon_*)$ in the case of
$M_i=M_{\rm BH}=10^{9}M_\odot$ and $10^{11}M_\odot$ respectively
in \autoref{fig:parameterspace}.


\subsection{Analysis of the BH mass-stellar mass relation}

The hierarchical Bayesian analysis that allows us to go beyond individual events
in order to study population properties (the $M_{\rm BH}-M_*$ relation)
is utilized in the analysis of LRDs.

In our model, the star formation efficiency, $\epsilon_*$, is
regarded as a redshift-independent constant. However, in reality,
one should consider the impact of black hole accretion, as well as
feedback from both black holes and TypeII Supernova
\citep{2024arXiv240707162D}. Therefore, $\epsilon_*$ should be
understood as the redshift-averaged star formation efficiency. As
mentioned, the stellar mass $M_*(z)$ of a galaxy seeded by an
SMPBH can be straightforwardly obtained using \autoref{eq:Mstar},
we denote it as $F(z, M_{\rm BH};z_{\rm cut},\epsilon_*)$,
however, due to various potential uncertainties in the model, we
assume that the probability distribution of the stellar mass $M_*$
at redshift $z$ for a given SMBH mass $M_{\rm BH}$ follows a
log-normal distribution parametrized by the mean $F(z,M_{\rm
BH};z_{\rm cut},\epsilon_*)$ and the width $\sigma$:
\begin{equation}
    \frac{{\rm d}P(M_*|z,M_{\rm BH},\Lambda)}{{\rm d}\log M_*}
    =\mathcal{N}\left(\left.\log M_*\right|F(z,M_{\rm BH};z_{\rm cut},\epsilon_*),\sigma\right),
\end{equation}
where the set of hyper-parameters $\Lambda=(\sigma,z_{\rm cut},\epsilon_*)$
describes the galaxy population seeded by SMPBHs and $\mathcal{N}(x|\bar{x},\sigma)$
denotes the probability density of a Gaussian distribution with mean $\bar{x}$
and variance $\sigma^2$.

In our Markov chain Monte Carlo analysis, $\theta\equiv\{z, M_{\rm BH}, M_*\}$
constitutes the parameters that define the individual LRD event.
We fit the parameters $\sigma, z_{\rm cut}$ and $\epsilon_*$ to
the measured data. The likelihood function is
\begin{align}
    \mathcal{L}(\boldsymbol{d}|\Lambda)&\propto\prod_j\int {\rm d}\log M_*
    ~{\rm d} z
    ~{\rm d}\log M_{\rm BH}\frac{{\rm d}P(M_*|z,M_{\rm BH},\Lambda)}{{\rm d}
    \log M_*} \notag \\
    &\times\mathcal{N}(\log M_*|\log M_{*,j},\sigma_{*,j})
    \mathcal{N}(z|z_{j},\sigma_{z,j}) \notag \\
    &\times\mathcal{N}(\log M_{\rm BH}|\log M_{{\rm BH},j},\sigma_{{\rm BH},j}),
\end{align}
where the product is over the data points, $M_{*,j}$, $z_j$ and $M_{{\rm BH},j}$
denote the mean measured stellar masses, redshifts and SMBH masses, and we assume
that the posteriors of the stellar mass, redshift and SMBH mass measurements are
log-normal and uncorrelated. 


\section{Results and discussion} \label{sec:results}

\begin{figure}
    \centering
    \includegraphics[width=\linewidth]{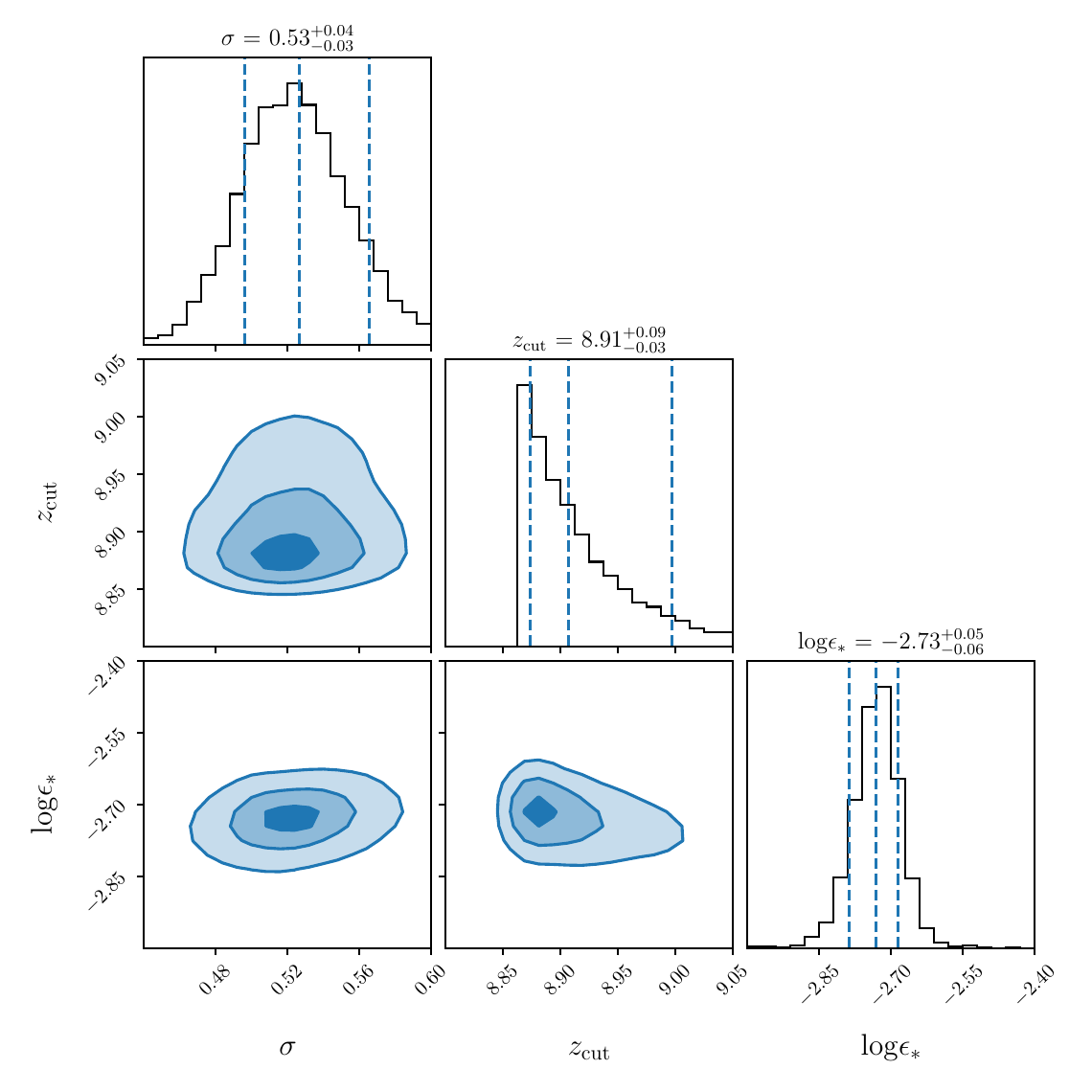}
    \caption{\label{fig:mcmc} The posterior distribution of the
    Bayesian parameter estimation. The contours in the 2D distribution
    represent $1\sigma$, $2\sigma$ and $3\sigma$ regions respectively.
    }
\end{figure}

In our analysis, we employ uniform priors for each parameter:
$\sigma$ in the range $[0,4]$, $z_{\rm cut}$ in the range
$[4,20]$, and $\log \epsilon_*$ in the range $[-10,0]$. We report
the Bayesian posterior distribution of parameters in
\autoref{fig:mcmc}. The median value of the parameters together
with their $90\%$ equal-tailed credible intervals are
$\sigma=0.53_{-0.03}^{+0.04}$,
$z_{\rm cut}=8.91_{-0.03}^{+0.09}$ and $\log \epsilon_*=-2.73_{-0.06}^{+0.05}$.
It is noted that the
cut-off redshift $z_{\rm cut}$ is constrained to be $z_{\rm
cut}>8.87$. This is because star formation is assumed to occur
only at $z<z_{\rm cut}$, and the maximum redshift of the LRDs in
our analysis is approximately $\sim8.86$ (see the left panel of
\autoref{fig:massrelation}). And the required star formation
efficiency is fully compatible with the results estimated in
\citet{2021ApJ...922...29S}.

In this work, we present a novel explanation for the anomalies
associated with LRDs by invoking SMPBHs. Unlike conventional
astrophysical seed mechanisms, which often require sustained
(super-)Eddington accretion, SMPBHs that came into being during
the radiation-dominated era offer a more natural and efficient
pathway to the origin of the massive SMBHs observed in
high-redshift galaxies. Additionally, SMPBHs can act as catalysts
for early structure formation through the seed effect, providing a
straightforward explanation for the unexpectedly high number
density of LRDs without the need for intricate astrophysical
processes. Based on a simple model, we demonstrates that galaxies
seeded by SMPBHs can readily attain the elevated black
hole-to-stellar mass ratios recently revealed in the COSMOS-Web
survey \citep{2024arXiv240610341A}. We further show that the
extreme masses of these SMPBH-seeded galaxies imply that only a
low star formation efficiency ($\log \epsilon_*\sim-2.73$) is
required to account for their stellar mass.

A larger sample of red and compact objects at high redshift,
coupled with improved understanding of the uncertainties in their
black hole and stellar mass measurements, is crucial for
confirming the existence of this extraordinary population of
overmassive black holes and their broader cosmological
implications. We have shown that the hypothesis that these
high-redshift massive galaxies are seeded by SMPBHs is supported
by several considerations. However, our analysis focuses solely on
the SMPBH scenario and it is important to explore a quantitative
comparison with intermediate-mass PBH models
\citep[e.g.][]{2024arXiv240707162D} and astrophysical models
\citep[e.g.][]{2024arXiv241006141P}.
The deeper observations across X-ray, mid/far-IR, and radio
wavelengths will be essential to accurately depict the intrinsic
properties of LRDs \citep{2024arXiv240610341A}, which helps us to
distinguish the origin of SMBHs. The relevant issues will be left
for the future to investigated


A refined SMPBH model that incorporates numerical simulations
\citep[e.g.][]{2024MNRAS.528..180Z,2024arXiv240218773J,2024arXiv240511381Z}
and accounts for more realistic astrophysical environments will also be important
for advancing our understanding. Such studies could offer critical insights
into the role of PBHs in early structure formation and help clarify the
mechanisms responsible for producing these SMBHs in the
high-redshift ($z\gtrsim4$) universe.


\section*{Data Availability}

The data underlying this article will be shared upon request to the
corresponding author(s).

\section*{Software}

This research made use of \texttt{numpy} \citep[\url{https://numpy.org}]{numpy}, \texttt{matplotlib} \citep[\url{https://matplotlib.org/}]{matplotlib} and \texttt{emcee} \citep[\url{https://github.com/dfm/emcee/}]{emcee}.

\section*{Acknowledgements}

We would like to thank Dashuang Ye for useful discussions. We are
also thankful for the observations made with the NASA/ESA/CSA
James Webb Space Telescope and the public data available in
\url{https://github.com/hollisakins/akins24_cw}. This work is
supported by National Key Research and Development Program of
China (Grant No. 2021YFC2203004), NSFC (Grant No.12075246), and
the Fundamental Research Funds for the Central Universities.




\bibliographystyle{mnras}
\bibliography{astro} 




\begin{appendices}

\section{Compatibility of the SMPBH abundance with current observations}
\label{appendixB}

It is well-known that the abundance of SMPBHs is strictly limited
by cosmological and astrophysical observations. In this Appendix,
we will check whether the required abundance of SMPBHs to explain
the energy density of SMBHs is compatible with current
observations.

Recently, many collaborations have constructed the SMBH mass
function, $\Phi(M_{\rm BH})$, using different sample
\citep[e.g.][]{2024arXiv240906772T,2024ApJ...963..129M,2024ApJ...962..152H,
2022MNRAS.517.2659W,2012ApJ...746..169S}. To achieve this, they
divide the SMBH mass range into several bins with widths denoted
as $\Delta\log \left(\frac{M_{\rm BH}}{M_\odot}\right)$ (Here for
convenience we will refer to it as $\Delta\log M_{\rm BH}$ and
this convention has been used throughout the text). Then each SMBH
is fractionally assigned to these mass bins according to its
posterior distribution. The mass function is constructed by
dividing the weighted object count in each bin by the bin width
and the comoving volume. An example of the result can be seen in
the Figure 12 of \citet{2024arXiv240906772T}.

Assuming that all the SMBHs are of primordial origin, we can
convert their mass function into the fraction of DM,
\begin{equation}
    f_{\rm PBH}(M_{\rm PBH})\equiv\frac{\Omega_{\rm PBH}(M_{\rm PBH})}{\Omega_{\rm DM}}=
    \Delta\log M_{\rm BH}\times\frac{\Phi(M_{\rm BH})M_{\rm BH}}{\rho_{\rm DM}}.
\end{equation}
We compare the results with current observational constraints on
$f_{\rm PBH}$ in \autoref{fig:smpbh}. It is obvious that the
required abundance of SMPBHs to explain the energy density of
SMBHs is consistent with existing upper limits.

\section{Analytical solution for halo mass growth} \label{appendixA}

The solution of the ordinary differential equation
\begin{equation}
    \frac{{\rm d}y(x)}{{\rm d}x}=[y(x)]^{\nu}f(x)
\end{equation}
is given by
\begin{equation}
    y(x)=\left\{[y(x_0)]^{1-\nu}-(-1+\nu)\int_{x_0}^{x}f(x')dx'\right\}^{\frac{1}{1-\nu}}.
\end{equation}
According to \autoref{eq:dotMh}, the mass growth rate of DM halos
after $z_{\rm cut}$ is given by
\begin{equation}
    \frac{\rm d}{{\rm d}z}\left(\frac{M_h}{M_\odot}\right)\simeq\left(\frac{M_h}{M_\odot}\right)^{1.1}f(z),
\end{equation}
where
\begin{equation}
    f(z)=-0.042\times\frac{1+1.11z}{1+z}.
\end{equation}
We use the fact that
\begin{equation}
    \frac{{\rm d}t}{{\rm d}z}=-\frac{1}{H_0\sqrt{\Omega_m(1+z)^3+\Omega_\Lambda}}
    \frac{1}{1+z},
\end{equation}
where $H_0$ is the Hubble constant. Throughout this paper our
cosmological parameters are set in the corresponding Planck values
\citep{2020A&A...641A...6P}, and we do not consider the effect of
the Hubble tension on $H_0$ and relevant results \citep[e.g.][]{2020PhRvD.101h3507Y,
2024arXiv240418579W},
which may be actually negligible. Therefore, the analytical
solutions of \autoref{eq:dotMh} is
\begin{align}
    \frac{M_h(z)}{M_\odot}=&\left\{\left[\frac{M_h(z_{\rm cut})}{M_\odot}\right]^{-0.1}
    -0.004662(z_{\rm cut}-z)+0.000462\right. \notag \\
    &\left.\ln\left(\frac{1+z_{\rm cut}}{1+z}\right)\right\}^{-10}.
\end{align}

\end{appendices}

\bsp    
\label{lastpage}
\end{document}